\documentclass[floatfix,lengthcheck,showpacs,amssymb,amsmath,amsfonts,twocolumn,nofootinbib,longbibliography]{revtex4-1}
\usepackage{amsfonts,amsmath,units,wasysym,epsfig,graphicx,verbatim,color,subfigure,graphicx}
\usepackage{amsmath}
\usepackage{latexsym}
\usepackage{amssymb}
\usepackage{amsfonts}
\usepackage{mathtools}
\usepackage{bm}
\usepackage{color}
\usepackage{float}
\usepackage{tikz}
\usepackage{adjustbox}
\usepackage{array}
\usepackage{soul}
\usepackage{appendix}
\usepackage{physics}
\usepackage{braket}
\usepackage{xcolor}
\usepackage[none]{hyphenat}

\usepackage{lineno}

\newcommand{\Real}{\operatorname{Re}}

\def\be{\begin{equation}}
\def\ee{\end{equation}}
\def\bea{\begin{eqnarray}}
\def\eea{\end{eqnarray}}

\begin{document}
\author{Rahul Dhurkunde}
\author{Henning Fehrmann}
\author{Alexander H. Nitz}
\affiliation{Max-Planck-Institut fur Gravitationsphysik (Albert-Einstein-Institut), D-30167 Hannover, Germany}
\affiliation{Leibniz Universitat Hannover, D-30167 Hannover, Germany}

\date{\today}

\title{A hierarchical approach to matched filtering using a reduced basis}
\begin{abstract}

Searching for gravitational waves from compact binary coalescences (CBC) is performed by matched filtering the observed strain data from gravitational-wave observatories against a discrete set of waveform \textit{templates} designed to accurately approximate the expected gravitational-wave signal, and are chosen to efficiently cover a target search region. The computational cost of matched filtering scales with both the number of templates required to cover a parameter space and the in-band duration of the waveform. Both of these factors increase in difficulty as the current observatories improve in sensitivity, especially at low frequencies, and may pose challenges for third-generation observatories. Reducing the cost of matched filtering would make searches of future detector's data more tractable. In addition, it would be easier to conduct searches that incorporate the effects of eccentricity, precession or target light sources (e.g. subsolar). We present a hierarchical scheme based on a reduced basis method to decrease the computational cost of conducting a matched-filter based search. 
Compared to the current methods,  we estimate without any loss in sensitivity, a speedup by a factor of $\sim$ 10 for sources with signal-to-noise ratio (SNR) of at least $= 6.0$, and a factor of $\sim 6$ for SNR of at least 5. Our method is dominated by linear operations which are highly parallelizable. Therefore, we implement our algorithm using graphical processing units (GPUs) and evaluate commercially motivated metrics to demonstrate the efficiency of GPUs in CBC searches. Our scheme can be extended to generic CBC searches and allows for efficient matched filtering using GPUs.
\end{abstract}
    \maketitle

\section{Introduction}
The first gravitational wave (GW) detection in 2015 marked the dawn of GW astronomy \cite{gw150914}. The first two observation runs of LIGO \cite{LIGOdetector} and VIRGO ~\cite{VIRGOdetector} detectors (O1 and O2) reported over a dozen confident detections~\cite{GWTC1, 2ogc}. The number of detections has rapidly increased to over 50 with the most recent O3 observing run\cite{GWTC2,Nitz:2021uxj}. To date, all gravitational-wave observations have come from compact binary coalescences (CBC); the vast majority of sources were from binary black holes (BBH) ~\cite{GWTC2, Nitz:2021uxj}, but notably two binary neutron star mergers (NS) ~\cite{BNS-1, BNS-2}, and recently two neutron star -- black hole NSBH mergers \cite{NS_BH} have been observed. These observations have helped us to understand the physics of compact objects \cite{physics_NS, NS_eq_state} and their dynamical evolution \cite{physics_NS}. 
As the gravitational-wave observatories become more sensitive, the increased number of CBC sources will allow us to determine merger rate \cite{event-rate} and population distribution \cite{BBH_pop}. Upcoming third-generation observatories such as the Einstein telescope \cite{ET}, cosmic explorer \cite{CE-intro}, and LISA \cite{LISA-intro} are expected to detect new kinds of astrophysical sources \cite{ET, CE-newsources, LISA-newsources, evans2021horizon}. 

\textit{Matched filtering} is the most widely used technique to detect CBC signals ~\cite{gstlal_offline, mbta_latest, 2ogc, SPIIR_GPU}. The method is optimal for stationary Gaussian noise~\cite{Creighton-book}. While the detector data contains non-Gaussian noise transients ~\cite{Glitches1, Glitches2}, which require the use of vetoing techniques \cite{bruce_chisq, Vetoes1}, matched filtering remains the dominant computational cost of a search algorithm ~\cite{find_chirp}. In this work, we focus only on the implementation of matched filtering. Matched filtering requires accurate models of the expected gravitational waveform; CBCs can be modeled using different techniques \cite{Phenom, EOB, NR}. The parameters of a binary merger are categorized into intrinsic (e.g. masses and spins) and extrinsic (e.g. binary orientation and location). To search for sources with unknown intrinsic parameters, we must select a discrete bank of templates which span the parameter space. These templates are chosen such that the minimum match (MM) between the data and at least one template from the bank is sufficiently large \cite{TB_hybrid, TB_effectual}. For example, a minimum match value of $~\sim 0.97$ would imply that at least $97\%$ of the SNR of any signal with parameters within the search area could be recovered. To identify a potential signal, gravitational-wave strain data is convolved with every template in a bank to calculate a signal-to-noise time series. Extrinsic parameter are often analytically maximized over. A possible candidate is identified if the SNR rises above a pre-determined threshold, passes various tests of signal consistency~\cite{bruce_chisq, Vetoes1} and data quality \cite{Data_quality1OGC, Glitches1, Data_quality}, and is statistically significant ~\cite{GWTC1, 1OGC}. The computational cost of matched filtering, and so also the entire search, scales linearly with the number of templates in a bank and also increases with the duration of the observable signal, though generally sub-linearly. 

With advancements in the current and future detectors, it is expected that observation of signals at an increasingly lower frequency will become possible~\cite{ET_data_analysis, ET, LISA-newsources, CE-newsources, 3g_sensitivity}. As the low frequency cutoff of a search decreases, both the size of the template bank and the signal duration grows rapidly, leading to increased computational costs \cite{ET_data_analysis, Nitz_ecc_search, LIGO_subsolar}. While current template-based searches have confident detections only from quasi-circular aligned-spin binaries~\citep{gstlal_offline, 2ogc, GWTC2, mbta_latest}, sources that exhibit measurable eccentricity or precession of the orbital plane could provide unique astrophysical insights ~\cite{Prospects_eccentricity, Prospects_precession}. While a few searches have included the effect of eccentricity \cite{LIGO_ecc_search, Nitz_ecc_search, Lennon_ecc_search} for parts of parameter space, many searches neglect the effects from eccentricity of the orbital plane \cite{eccentricity} and precession of the orbit \cite{precession} in part due to the increased computational cost relative to normal searches~\cite{ianharry_precession, Nitz_ecc_search, Nitz:2021uxj, LIGO_subsolar}. For example, it has been shown that the template bank including precession is at least 10 times bigger than one without precession \cite{ianharry_precession}. 
Furthermore, the lower mass boundary of subsolar primordial black hole searches is limited by computational cost considerations ~\cite{LIGO_subsolar, ET_data_analysis, Nitz_ecc_search, Nitz:2021uxj}. Development of cost-efficient filtering algorithm would allow searches to be conducted more easily, with higher sensitivity, and in uncharted regions of parameter space. 

The computational costs for matched filtering data with templates consist of redundant computations due to significant overlap of templates with each other in the neighborhood. This redundancy is be eliminated by using an orthonormal basis to filter data instead of templates \cite{SVD_initial}. The costs of filtering scale linearly with the number of basis and can be reduced by rejecting the basis vectors of lower importance \cite{SVD_initial}. Disregarding contributions from a few basis vectors leads to a loss in SNR, but this loss is kept under a tolerance by tuning the number of relevant basis $p$
involved for filtering purpose \cite{SVD_initial}. When considering large number of templates, the value of $p$ is much smaller, and therefore, it is possible to filter data with a fewer number of basis. Current online (lwo-latency) searches \cite{GSTlal_online, SPIIR_GPU} employ this technique and are in good agreement with searches that do not use this approximation~\cite{pycbc_offline, mbta_online}.

Since the basis vectors do not correspond to any physical source, match filtering outputs from each basis are weighted and linearly combined to give an SNR time-series for a unique template, in a~\textit{reconstruction} process. The reconstruction is performed for each template and at every time sample which occurs additional costs to matched filtering. The naive costs for reconstructing SNR time-series for $T$ templates can be estimated in terms of a matrix multiplication which scales as $\mathcal{O}(NpT)$, where $p$ is the number of basis and $N$ the number of samples in the data. On the other hand, the direct template-based filtering is widely done using a Fast Fourier Transform (FFT) - based algorithm which requires $\mathcal{O}(NT\log{N})$ operations. Usually $\log N \sim O(10)$ and $p \sim O(10^2-10^3)$ which suggests that naive reduced basis filtering is more expensive than the template-based filtering. 

In this work, we demonstrate a cost-efficient matched filtering method by employing a new hierarchical scheme using a reduced set of basis. The reduced basis are obtained by applying principal component analysis (PCA) on a template bank. A two-stage hierarchical scheme is then invoked to compute the SNR time series for each template. In the first stage, an intermediate time-series is computed that corresponds to a binned average of the complete SNR time-series. In the second step we, do a full time resolution (non-averaged) reconstruction using the reduced basis inside the bins where the average SNR exceeds a threshold. We demonstrate our method on simulated Gaussian noise and a population of CBC signals. To estimate the improvement from our method, we compare it against the flat template based filtering scheme used in current searches. We observe that our method attains a speed up by a factor of $\sim \textit{6}$ for a threshold of SNR = 5. Furthermore, we expect the performance of our method to increase at higher SNR thresholds, and similarly observe a performance gain of $\sim \textit{10}$ times for SNR = 6. 

Matched filtering is dominated by mathematical operations which are easily parallelizable across different threads or compute cores. Since a GPU is designed to accommodate a large number of threads, we employ GPUs in this work for an efficient implementation of matched filtering in parallel. To investigate the relative performance of different hardware, we compare the matched filtering implementations on GPUs with the central processing units (CPUs) that is currently used in the PyCBC search pipelines. We use two metrics to quantify the performance -- cost, and power efficiency while filtering data. We observe that GPUs are much more efficient than CPUs in performing matched filtering. In this analysis, we have restricted ourselves to aligned-spins, dominant mode, and a single detector analysis. However, our method can be easily extended to multiple detectors, and search scenarios including eccentricity, precession, or higher-order modes.

The rest of the paper is organized as follows. In section \ref{Sec:CBC_searches}, we give a brief overview of the matched filtering process and motivate why we need a new approach different from previous works. In section \ref{Sec:Search_algorithm} we describe our search method and the implementation. In section \ref{Sec:Results} we presents our results and compare them with the existing search methods. Finally we conclude in section \ref{Discussion_and_future_prospects}.

\section{Compact binary coalescing searches} \label{Sec:CBC_searches}

The core of any modeled searches for CBC signals is the matched filtering technique which involves searching the interferometric data for a modeled waveform of the expected GW signal \cite{2ogc, gstlal_offline, SPIIR_GPU, mbta_latest}. In this section we briefly describe the process of matched filtering and introduce some ideas for efficient filtering algorithms. We also discuss previous efforts to improve the performance of matched filtering in section \ref{Sec:Current-methods}. 

GW signals from a non-eccentric CBC sources are characterised by 15 parameters \cite{Phenom, EOB, NR}. These parameters are divided in two categories, 1) Intrinsic parameters - ($m_1, m_2$) component masses, and three dimensional spin vectors ($\vec{\chi_1}, \vec{\chi_2}$). 2) Extrinsic parameters (in the observer frame of reference) - standard spherical coordinates ($D, i, \psi$), sky-location ($\theta, \phi$), and lastly ($t_c, \phi_c$) the time and the phase at the coalescence. The anticipated signal is then accurately modeled in terms of these parameters with the help of various analytical and numerical techniques \cite{Phenom, NR, EOB}.

To search for the modeled signal $\tilde{h}(f)$ also referred as template, matched filtering is performed in the Fourier domain to quantify the likelihood of data containing the particular template. The matched filter statistic is a correlation between the Fourier transform of the data ($\tilde{s}(f)$) and the template ($\tilde{h}(f)$) weighted by the noise power spectral density (PSD) $S_n(f)$ \cite{Creighton-book}. It can be shown that matched filter is an optimal detection statistic for distinguishing signals in the presence of stationary Gaussian noise \cite{Creighton-book}. The mathematical form of the complex matched filter statistic is 
\begin{equation}
\label{MF_def}
    \langle s|h \rangle = 4\int\limits_{0}^{\infty}\dfrac{\tilde{s}(f)\tilde{h}^*(f)}{S_n(f)} df.
\end{equation}
The output of the matched filter after normalizing with the correlation of the template with itself $\Braket{h|h}^{1/2}$ is the signal-to-noise ratio SNR
\begin{align}
    \label{SNR_def}
    \rho^2 = \dfrac{\big(\Real{[\Braket{s|h}]}\big)^2}{\Braket{h|h}}.
\end{align}

Apriori the parameters of gravitational-wave signals are unknown and to search for the intrinsic parameters, a discrete template bank is used to \textit{cover} the intrinsic parameter space.  The notion of cover is to sample enough points in the parameter space such that the match between data and at least one template is above a minimum match value. In current searches typically a minimum match of 0.97 is used \cite{Searching_CBC, find_chirp}, and lattice-based \cite{tau0_tau3, Searching_CBC, TB_lattice1}, stochastic methods \cite{TB_stochastic1, TB_stochastic2} or hybrid methods \cite{TB_hybrid, TB_effectual} are applied to sample the points in the parameter space. Since we are considering aligned-spins with the orbital angular momentum in the $+z$ direction, the intrinsic parameter space consists only of $\zeta = (m_1, m_2, \chi_{1z}, \chi_{2z})$ parameters. Search over the two categories of binary parameters are handled differently - the intrinsic parameters are searched by repeatedly matched filtering for every template. Whereas, the extrinsic parameters - sky location, the orientation of the binary, and distance to the source are accounted as an overall phase $\phi_0$ and an amplitude $A$ \cite{find_chirp}

\begin{align}
    \label{hf_eqn}
    \tilde{h} = Ae^{i\phi_0}\tilde{h}_0( \zeta) e^{2 \pi i f t_c}.
\end{align}

In the Eq. (\ref{hf_eqn}), $A$ and $\phi_0$ are unknown functions of $(D, i, \theta, \phi, \psi)$, and $h_0$ depends on the intrinsic parameters. The unknown amplitude $A$ is a nuisance parameter that is eliminated by normalizing the SNR with the norm of the template as seen in Eq. (\ref{SNR_def}). The unknown phase $\phi_0$ is maximised using a quadrature, which is equivalent to maximizing the norm of the complex SNR \cite{find_chirp}. Finally, the position of the signal is determined by searching for the time of coalescence of binary, represented by the $t_c$ parameter. Variation in $t_c$ is expressed as time-translations, and is separated using $e^{2\pi i f t_0}$. Substituting Eq. (\ref{hf_eqn}) in Eq. (\ref{MF_def}), the matched filter output (SNR) at $t = t_0$ is given by Eq. (\ref{IFFT}). The SNR as a function of time can be obtained efficiently by performing an inverse FFT (IFFT) of the Eq. (\ref{IFFT}) \cite{find_chirp, tau0_tau3}
\begin{align}
    \label{IFFT}
    \Braket{s|h_0}(t_0)  & = 4\int\limits_{0}^{\infty}\dfrac{\tilde{s}(f)\tilde{h}_0^*(f;\zeta)}{S_n(f)} e^{2\pi i ft_0}df.
\end{align}

Data segment with SNR above a predetermined threshold is referred as a \textit{trigger} which may contain a true GW signal. The ambiguity is due to the assumption of stationary Gaussian noise for the matched filter statistic. However, triggers due to non-stationary glitches or pure Gaussian noise can give rise to false alarms which lower our confidence of identifying true GW signals \cite{Glitches1, Glitches2}. Additional signal consistency test introduced in \cite{bruce_chisq, Vetoes1, Veto_unichisq} is performed to down-rank triggers due to glitches. Furthermore, it is ensured that only coincident triggers from multiple detectors are considered -- triggers corresponding to the same template and observed within the light travel time window between the detectors \cite{Usman_pycbc}. Amongst the various steps mentioned, matched filtering comprises the dominant computational costs of a search. Hence, our focus is to optimize the matched filtering process.  

To summarize the matched filtering procedure, the intrinsic and extrinsic parameters are searched separately using a template bank, analytical techniques respectively. First, the PSD weighted correlation of data with a single template in $\zeta$ is computed, and then an inverse FFT is executed to obtain a complex time-series. Taking the modulus of the complex times-series and normalizing it by the norm of the template gives the SNR times-series. The above steps are repeated for all the templates in the template bank to search over the intrinsic parameters. Throughout this paper, we refer to the method of matched filtering data with the templates as the \textit{template method} for simplicity.

It is clear from above that matched filtering operation scales linearly with the number of templates. In the case when the size of the template bank is large, a search can be limited by the computational costs required for filtering the data with templates. It is possible to numerically reduce the size of the template banks to a fewer number of basis vectors and filter data directly with the basis. We now give a brief introduction to performing matched filtering with a reduced basis. 

\subsection{Matched Filtering using a Reduced Basis} \label{Red_basis_theory}

Consider a region in the parameter space described by $\zeta = (m_1, m_2, \chi_{1z}, \chi_{2z})$. Discrete templates are used to cover this region, and as a result of the mismatch criterion, templates are strongly correlated in the vicinity of each other. The correlation between the templates incurs a redundancy in matched filtering computations, instead, an orthonormal basis can be used to eliminate these correlated computations \cite{SVD_initial}. Commonly used methods for computing an orthonormal basis is the principal component analysis (PCA) and the singular value decomposition (SVD). The PCA approach to obtain basis is by performing an eigenvalue decomposition (EVD) of the covariance matrix $\textbf{C} = \textbf{T}^{\top}\textbf{T}$ constructed using the templates (see Eq. (\ref{Eq-PCA})). Whereas, SVD is applied directly to a matrix containing the templates $\textbf{T}$ (see Eq.(\ref{Eq-SVD})). 
\begin{subequations}
  \begin{equation}
    \label{Eq-PCA}
      \textbf{C} = \textbf{PL}\textbf{P}^{\top},
  \end{equation}
  \begin{equation}
    \label{Eq-SVD}
    \textbf{T} = \textbf{US}\textbf{P}^{\top}.
  \end{equation}
\end{subequations}

In the case when templates are centered -- the column means of $\textbf{T}$ are zero, then both SVD and PCA yield the same orthonormal basis. The basis are represented as columns of $\textbf{P}$ which are ranked by their corresponding eigenvalues in $\textbf{L}$ or singular values in $\textbf{S}$. It can be easily shown that $\textbf{L} = \textbf{S}^2$ and that the two methods are similar. Hence, either of the methods are applicable to obtain the basis. 

Consider a set of $p_t$ basis vectors for the parameter region $\zeta$ denoted by $\tilde{p}(f)$ in the Fourier domain. Every template in this region $h_{\zeta}$ is expressed in terms of a unique linear combination of the basis. Since the matched filtering operation is also linear, we can filter the data $\tilde{s}$ only using the basis and rewrite Eq. (\ref{IFFT}) in terms of $\tilde{p}(f)$
\begin{align}
        \rho_r(t) &=  \sum \limits_{k=0}^{p_t-1} 4 \int\limits_{0}^{\infty}\dfrac{\tilde{s}(f)c_{k, \zeta}^*\tilde{p}_{k}^*(f)}{S_n(f)} e^{2\pi i ft}df, \label{basis_SNR}
\end{align}
where the template $\tilde{h}_{\zeta}(f) = \sum \limits_{k=0}^{p_t-1}c_{k, \zeta}\tilde{p}_{k}$, is a linear combination of the basis $\tilde{p}(f)$ and the unique decomposition coefficients $c_{k, \zeta}$. The coefficient $c_{k, \zeta}$ is obtained by computing the scalar product of $\tilde{h}_{\zeta}$ and the $k^{th}$ basis vector $\tilde{p}_{k}$. 

Matched filtering with the basis is done by first performing an IFFT of the correlation between $\tilde{s}$ and $\tilde{p}_k$, which results in a complex time-series defined as $\beta_{k}(t) = \text{IFFT} (\langle s|p_{k} \rangle)$. Afterwards, each $\beta_k$ is weighed accordingly using the respective decomposition coefficients and then combined to give SNR time-series corresponding to the template $\tilde{h}_{\zeta}(f)$. The process of multiplying the coefficients $c_{k,\zeta}$ with $\beta_{k}$ is the reconstruction step, as it reconstructs the SNR time-series using the contribution from the basis
\begin{align}
    \rho_r(t) =  \sum \limits_{k=0}^{p_t-1} c^*_{k, \zeta} \beta_{k}. \label{Beta_eqn}
\end{align}

Using the complete orthonormal basis $p_t = T$ where $T$ is the number of templates, Eq. (\ref{Beta_eqn}) reproduces exact results as Eq. (\ref{IFFT}). 

Instead of using the full basis, it is possible to approximately reconstruct the SNR with fewer basis vectors. Eigenvalues $\sigma$ are arranged in decreasing order and only the first $p$ basis vectors are chosen and the rest are discarded. It can be shown that the first $p$ basis vectors span an approximate lower rank subspace of the original parameter space. Neglecting contribution from some basis vectors leads to an average loss in SNR, which is shown to be a function proportional to the eigenvalues \cite{SVD_initial}
\begin{align}
    \label{SNR_loss}
    \Braket{\dfrac{\delta\rho}{\rho}} = 1 - \dfrac{\left|\sum \limits_{k=0}^{p-1}\sigma^2_{k}\right|}{\left|\sum \limits_{k=0}^{p_t-1}\sigma^2_{k}\right|}.
\end{align}
The equation above indicates that the number of relevant basis $p$ can be fine-tuned based on the choice of tolerance in loss of SNR. For detection purposes, we want to keep this loss under the mismatch value ($0.03$) due to the discreetness of the template bank. 

Reconstruction of SNR time-series is performed for every template and thus, the reduced basis approach requires additional costs to matched filtering. To estimate the reconstruction costs in brief (exact costs are estimated later in this paper), consider a data segment having $N$ samples, $T$ number of templates, and reduced $p$ basis vectors. The number of operations required for the reconstruction step is $O(NpT)$. Meanwhile, comparing the costs for template based filtering; which requires $O(NT\log N)$, and thus, the actual comparison boils down to $\log N $ and $p$. The exact values for $\log N$ and $p$ vary over the parameter region, but considering a ballpark it is observed that $p$ is typically 1-2 order in magnitude bigger than $\log N$. Hence, the reduced basis approach loses all the computational advantage of filtering against fewer basis.

\subsection{Comparison with Current Methods}\label{Sec:Current-methods}

Different methods in the past have been implemented to speed up the process of filtering either by reducing the latency of the search \cite{SPIIR_initial, mbta_latest, LLOID} or by decreasing the required computational costs \cite{bhooshan, hierarchy_mass, hierarchy_tau0}. For the former case, a reduced basis filtering technique along with multi-rate sampling is used with an intent to decrease the latency of matched filtering. And in the latter case, the aim is to reduce the filtering costs by using a multi-stage hierarchical filtering method. Some of these techniques have been already implemented in the current search pipelines \cite{SPIIR_GPU, GSTlal_online, mbta_online}, and are deployed in different search scenarios. We briefly discuss these various strategies and contrast our methodology next.

\subsubsection{Reduced basis filtering with or without multi-rate sampling}\label{Sec:RB-and-MRF}

To discretize a continuous signal, the Nyquist-Shannon criterion \cite{Nyquist} determines the sampling rate to be at least twice the highest resolvable frequency ($1/dt \geq 2f_{max}$) of the anticipated signal. This gives the relation for the number of samples in the data $N = (df dt)^{-1}$ where $1/dt, 1/df$ are sampling rate and sampling frequency respectively, suggesting the filtering costs increase when searching for higher frequencies. Because the frequency evolution of these signals is \textit{chirp} like; rapidly increasing towards the merger, this allows a low sampling rate at the earlier times and can be increased subsequently as the signal evolves. Using multiple sampling rates the matched filtering costs are reduced significantly \cite{LLOID, mbta_latest, SPIIR_initial}. 

Multi-rate sampling has been adopted by the MBTA \cite{mbta_latest} LLOID \cite{LLOID} and SPIIR \cite{SPIIR_initial} schemes that are implemented in current online pipelines for the prompt detection of signals \cite{SPIIR_GPU, GSTlal_online, mbta_latest}. The MBTA method performs matched filtering in the Fourier domain using the standard FFT approach to obtain SNR time-series. Whereas the LLOID and SPIIR methods perform time domain filtering by employing FIR or IIR filters respectively to compute an equivalent form of the matched filter Eq. (\ref{IFFT}). These filters are specially designed for whitening the data, a process which causes most latency in matched filtering \cite{GSTlal_online, mbta_latest, SPIIR_GPU}. The overall latency is further improved by using a reduced basis obtained by performing SVD of the IIR/FIR filters. Results from the online pipelines are very well in agreement with the rigorous offline searches \cite{pycbc_offline, gstlal_offline}, justifying the viability of multi-rate sampling and reduced basis filtering in CBC searches. 

The number of templates $T$ drastically increases when searching in sub-solar regions \cite{LIGO_subsolar, ET_data_analysis, Nitz_ecc_search, Nitz:2021uxj} or with additional parameters \cite{ianharry_precession, Nitz_ecc_search} e.g. eccentricity. In such search scenarios obtaining a reduced basis for the complete template bank is computationally limited. This is because SVD is performed on a template matrix whose size scales linearly with $T$ and might require infeasible amounts of memory for the template matrix. To address this issue we choose the PCA approach of computing the basis, which is performed on a covariance matrix whose size is independent of $T$, and therefore, making it feasible to obtain orthonormal basis even for large template banks.  

The major drawback of a reduced basis filtering is the large reconstruction costs, and hence, this method is avoided in extensive offline searches where the computational costs play an important role. In \cite{RP-1, RP-2} the authors have introduced a new technique to decrease the reconstruction costs by using Random Projections (RP) based reduced basis filtering method. Another approach to reduce the total costs is to split the matched filtering into multiple stages in a hierarchical fashion. Next, we discuss established hierarchical methods and compare them with our new hierarchical scheme.

\subsubsection{Hierarchical methods}

The crux of any hierarchical search is to perform a coarse and a fine search over the parameters involved in the hierarchy. In the past, the work in \cite{hierarchy_mass} proposed a two-stage hierarchical filtering on just a single chirp mass parameter, and the same work was extended in \cite{hierarchy_tau0} for three parameters - the component masses and the time of coalescence. The most recent works \cite{bhooshan, kanchan} in hierarchical approach to matched filtering extended the scheme to multiple detectors analysis. For the first stage data is down-sampled at 512 Hz, and filtered using a coarse bank of MM = 0.9. In the second stage, data is sampled at the full rate of 2048 or 4096 Hz, and filtered with a fine bank having MM = 0.97. Their method achieved $\sim 20$x speed up in comparison to the one-step search on simulated data containing only Gaussian noise. The latter work in \cite{kanchan} hierarchically searched advanced LIGO's first two observing run and recovered all the events presented in the GWTC-1 catalog \cite{GWTC1}.

To assign significance of any event, it is important to estimate the noise \textit{background} which is the trigger distribution due to noise only events \cite{Usman_pycbc}. In the work \cite{bhooshan, kanchan}, the performance gain comes at the expense of a poor estimation of the background. Since they do not follow the noise triggers until the second stage, the true background is mimicked by scaling the first stage background. This leads to an improper estimation of the significance for an event.

In this method, we present a new hierarchical method aimed at reducing the reconstruction costs. We perform a two-stage hierarchical reconstruction of the SNR time-series for the complete bank. We follow up all the triggers for SNR $\geq 5$ till the second stage, and therefore, are able to accurately determine the original noise background. Furthermore, our method incurs no loss in the search sensitivity. We discuss the methodology in detail in the next section.   

\section{Reduced basis hierarchical matched filter}\label{Sec:Search_algorithm}
In this section, we describe our new hierarchical approach to reduced basis matched filtering in detail. We first briefly review the PCA method in general and discuss a non uniform sampling technique to reduce the costs of performing PCA. Then we explain how PCA is applied on a template bank and the hierarchical method of filtering data along with their implementation on GPUs. Finally, we estimate the matched filtering costs in detail for the template based method and the reduced basis hierarchical scheme to compare the relative gain in performance.

\subsection{PCA using Non-uniform Sampling}

We first briefly explain the PCA procedure applied on $n$ vectors denoted by $\textbf{v}$. Every vector is centered by subtracting the mean vector $\textbf{v}_s = \textbf{v} - \textbf{b}$, where $\textbf{b} = 1/n\sum_{i=0}^{n} v_i$ is the mean vector. To ensure each vectors get equal weight, they are normalized w.r.t. the inner product defined on the vector space. Using the normalized and centered vectors $\hat{\textbf{v}}_s$, a covariance matrix is created $\textbf{C} = \hat{\textbf{v}}_s^{\top}\hat{\textbf{v}}_s$. An EVD is performed to get the orthonormal basis vectors $\textbf{p}_t$ of the $\textbf{C}$ matrix, which are ranked by the corresponding eigenvalues $\sigma$. The basis vectors corresponding to small eigenvalues are discarded, and the resulting set of reduced basis is denoted by $\textbf{p}$.  Projecting $\hat{\textbf{v}}_s$ onto $\textbf{p}$ gives the decomposition coefficients $\textbf{D} = \textbf{p}\hat{\textbf{v}}_s$. The reduced basis $\hat{\textbf{v}}_s$ and the decomposition coefficients $\textbf{D}$ are used to retrieve an approximate version of $\hat{\textbf{v}}_s$, given by $\hat{\textbf{v}}_s^{approx} = \textbf{D}^{\top}\textbf{p}$. 

Even though the costs of PCA are amortized, PCA can be time-intensive and difficult to perform on a large collection of longer duration templates e.g. corresponding to lower masses, or with lower-frequency cutoffs. Amongst the various steps involved in PCA, the dominant computational and memory costs are for the covariance matrix -- both scale quadratically with the number of samples $N_t$ required for the templates. To put the scaling relations into perspective, consider the complete O2 bank sampled at a constant sampling frequency of 128s, the estimated memory required for $\textbf{C}$ is $\sim 550$ GBs. Distribution of the EVD process for $\textbf{C}$ across several machines is a difficult task, and thus, the size of the covariance matrix is constrained due to the memory of a single machine. Hence, it is crucial to reduce the size of the templates to make the PCA faster and feasible in the low-mass regime.

In this work, we consider the frequency range from [15, 1024] Hz. The principle idea behind efficient sampling is to adjust the sampling rate according to the number of oscillations of a function within a given frequency bin. To account for the complete frequency range in our sampling analysis, we consider the template with the longest bandwidth and identify all the frequencies corresponding to the zero crossings of this template. The identified frequencies are used to define edges of the non-overlapping bins of different sizes. We further sample every bin by using five uniformly spaced frequencies within, and together they make up the complete set of non-uniform sampling frequencies. We also ensure that our sampling criteria is never less than $df = 1/128$, which helps us avoid oversampling the dense bins at very low frequencies. The number of frequencies per bin is chosen empirically based on the relative error induced in the templates for not using the full sampling rate. Using this scheme we obtain a much smaller set of frequencies $N_t = 11074$ compared to $N=2048\times128$ uniform frequencies for efficient sampling of the templates and the basis. Once the basis are obtained, we interpolate them back to the original sampling frequencies.

To test the accuracy of our sampling method, we check the overlap between the templates generated using non-uniform and uniformly sampled frequencies. For this purpose, templates evaluated at non-uniform frequencies are linearly interpolated to a constant sampling frequency of 128s. We obtain a mismatch of $< 10^{-4}$ in the overlap due to interpolation of the templates, which is much smaller than the error due to discreteness of the template bank and, therefore, can be safely neglected. This justifies that our non-uniform frequencies are viable for sampling the templates. Using our sampling method significantly reduces the memory required for $\textbf{C}$ from $\sim 550$ GBs to only $\sim 0.9$ GBs. Therefore, saving a lot of computational resources and simultaneously speeding up the PCA process.

\subsection{Implementing PCA on a Template Bank}

In this work, we use the template bank as described in \cite{TB_used} which was also used for the PyCBC analysis of the O2 observing run \cite{2ogc}. The parameters ranges used in this bank are -- total mass $M \in [2, 100]$ and mass ratio $q \in [1, 98]$. We restrict ourselves to the aligned-spin case where the spins for NS are up to 0.05 and up to 0.998 for BHs. The minimum match criterion used in this bank is 0.97, and the bank contains $T_{total} \sim 400,000$ templates.

\begin{figure}[H]
    \centering
    \includegraphics[width=\linewidth, height = 6.5cm]{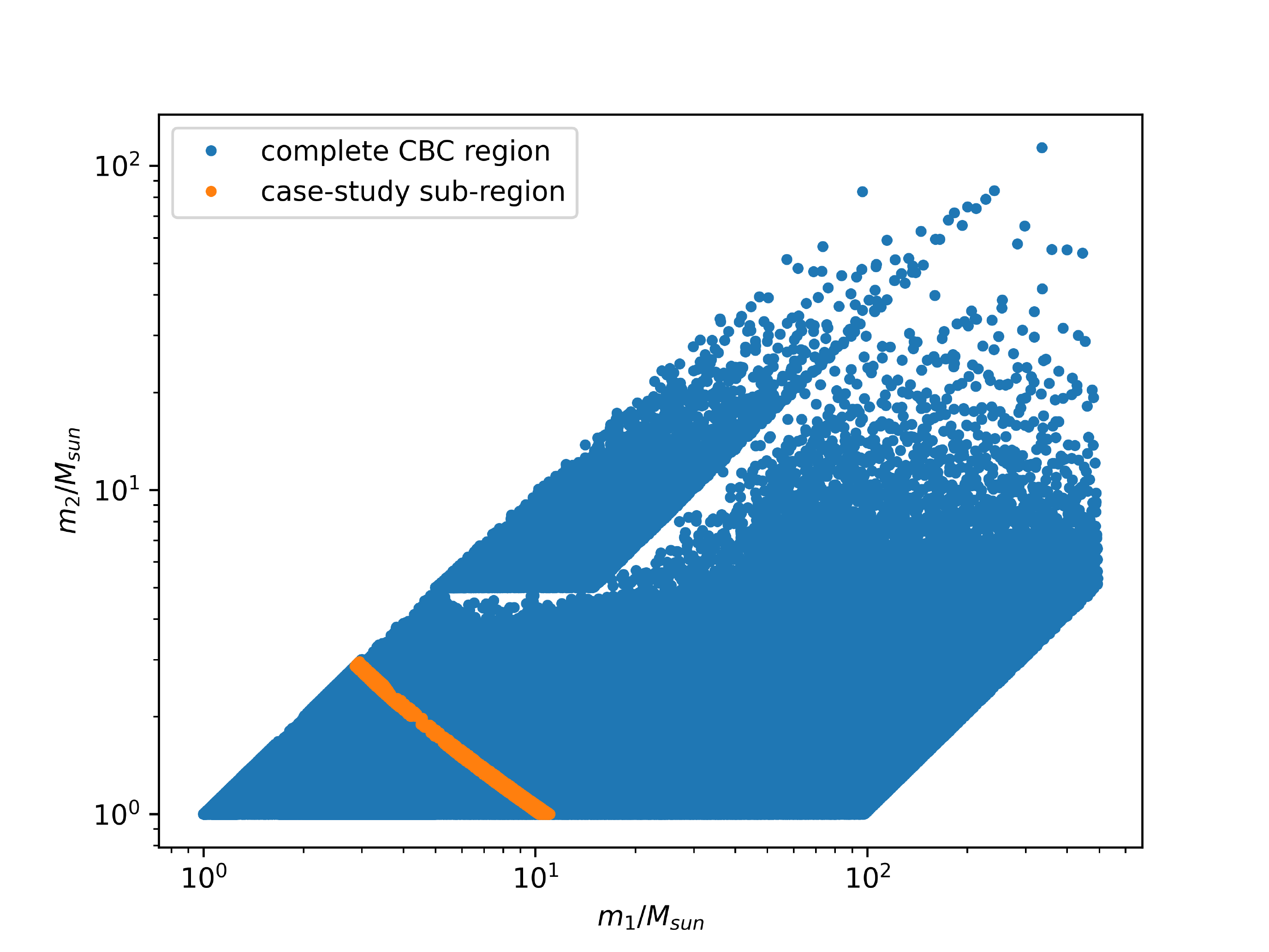}
    \caption{O2 template bank \cite{TB_used} used in this work, and the small orange region corresponds to the sub-bank used as a case study  containing 6250 templates.}
    \label{Bank}
\end{figure}

We divide the parameter space into smaller regions to reduce the number of local basis $p$ as they contribute linearly to the dominant reconstruction costs. The complete bank is split into smaller sub-banks, and then PCA is performed on each of them individually. Splitting of template bank is performed in the ($\tau_0, \tau_3$) coordinates along iso-$\tau_0$ lines because the metric is roughly Euclidean in these coordinates \cite{tau0_tau3}. The parameter $\tau_0$ roughly corresponds to the duration of a template in seconds, that scales as $\tau_0 \propto \mathcal{M}^{-5/3} f_{0}^{-8/3}$, where $\mathcal{M}$ is the chirp mass and $f_{0}$ is the lower frequency used in the analysis. We choose to split the $\tau_0$ range into 64 equal parts, each of them containing 6250 templates. While optimizing the splitting is not in the scope of this work, we performed empirical testing of the number of splits by considering smaller or bigger equal parts than 64, and observed no significant improvement. In Fig. \ref{Bank} we show the complete parameter space along with an example sub-region which is used as a case study for further analysis.

The PCA operation begins by sampling templates at the previously obtained $N_t$ non-uniform frequencies using the IMRPhenomPv2 waveform model \cite{Phenom}. Templates are then whitened using the aLIGO PSD and normalized to unity. The template matrix for the $m^{th}$ sub-bank $\textbf{T}^m$ is constructed by storing templates row-wise such that $\textbf{T}^m$ has the dimensions of ($T \times N_t$), where $T = T_{total}/64$. In case when $T_{total}$ is not a multiple of 64, we can simply choose another divisor close to 64. We observed the mean vector of $\textbf{T}^m$ to be almost zero, and hence, skip the mean subtraction step. Covariance matrices for each sub-bank $\textbf{C}^m$ are evaluated by multiplying the template matrix with its transpose $\textbf{T}^m \times (\textbf{T}^{m})^{\top}$. In the next step, we perform the EVD of $\textbf{C}^m$ to obtain the basis vectors and their corresponding eigenvalues. For this purpose we employ the Lanczos algorithm \cite{Lanczos} -- an efficient algorithm to obtain the $p$ largest eigenvalues. Invoking Eq. (\ref{SNR_loss}) for a tolerance of $10^{-5}$,  the number of relevant eigenvalues $p$ obtained for a few different sub-banks are shown in the Table \ref{table:sub-banks}.

\begin{table}[H]
    \begin{center}
    \begin{tabular}{|c|c|c|}
    \hline
     sub-bank & $\tau_0$ &$p$\\
     index &  (sec) & \\ 
     \hline
     1 & [0.1, 5.1] & 64 \\
     \hline
     . & . & . \\
     . & . & . \\
     \hline
    34(case study ) & [98.0, 103.4] & 254\\
     \hline
     . & . & . \\
     . & . & . \\
     \hline
     64 & [442.5, 595.7] & 200\\
     \hline
    \end{tabular}
    \end{center}
    \caption{Example of a few sub-banks corresponding to their respective $\tau_0$ ranges in the second column. The third column is the number of relevant eigenvalues $p$ for the respective sub-region.}
    \label{table:sub-banks}
\end{table}

We then compute the decomposition coefficients essential to reconstruct the whitened templates in $\textbf{T}^m$. These coefficients are unique for each template and are obtained by multiplying the two matrices - basis matrix $\textbf{P}^m$ and the template matrix $\textbf{T}^m$. The resulting matrix is the decomposition matrix $\textbf{D}^m$ containing the $p$ unique coefficients for every template in the $m^{th}$ sub-bank and has the dimensions (${p \times T}$). Finally, the original whitened templates can be approximately reconstructed by multiplying $\textbf{D}$ and $\textbf{P}$. In the next section, we discuss the two-stage hierarchical reconstruction of the matched filter output using $\textbf{D}$ and $\textbf{P}$.

\subsection{Hierarchical Reconstruction of the SNR Time-series}\label{sec:first_stage}
Matched filtering with the basis vectors as per Eq. (\ref{Beta_eqn}) is performed using the following steps:
\begin{itemize}
    \item compute FFT of the data $s(t)$ with $N$ sample points at a uniform sampling rate $1/dt$ to obtain $\tilde{s}(f)$.
    \item linearly interpolate the basis vectors $\tilde{p}_k(f)$ at the uniform frequencies (multiples of $2f_{max}/N$).
    \item filter data with every basis vector to obtain $\beta_k$ -- inverse FFT of the product $\tilde{s}(f) \tilde{p}^*_{k}(f)/\sqrt{S_n(f)}$ .
    \item average $\beta_k$ in bins of $w$ samples to obtain $\beta^{avg}_k$.
    \item perform first stage reconstruction to obtain averaged SNR time-series.
    \item perform second stage reconstruction around the triggers from the first stage.
\end{itemize}

The first step in the reduced basis matched filtering process is to compute forward FFT of the data $s(t)$. Since $s(t)$ is sampled uniformly with a rate of $1/dt = N df$, the Fourier transformed data $\tilde{s}(f)$ is obtained at frequencies given by integer multiples of $2f_{max}/N$, where $f_{max} = 1/(2dt)$. Now to filter the data with the basis, the correlation of data and basis are computed at the uniform frequencies, and for this reason, the basis are linearly interpolated from $f_{min}$ to $f_{max}$ in $df$ steps. Since the basis are already whitened and the denominator in the matched filtering Eq. (\ref{basis_SNR}) requires $S_n(f)$, we multiply the correlation product with $S_n(f)^{-1/2}$ to get the appropriate denominator. For a basis vector $\tilde{p}_k$, the filtered output $\beta_k$ time-series is obtained by computing the inverse FFT of the weighted correlation. Finally, for every data, $p$ different time-series (basis output) are stored in a separate $\pmb{\beta}$ matrix of size $N \times p$. 

\begin{figure}[ht]
    \centering
    \includegraphics[width=\linewidth, height = 8.0 cm]{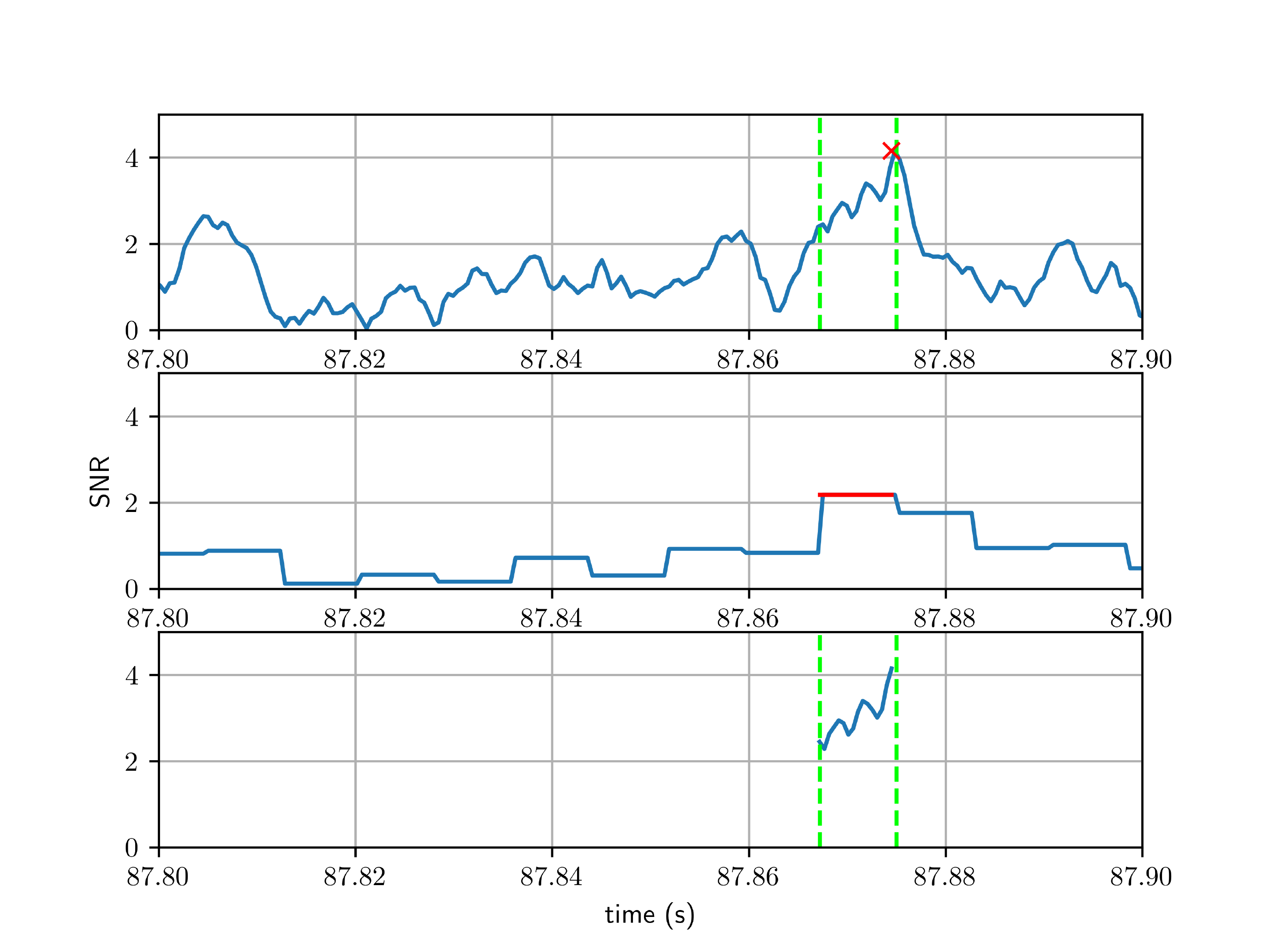}
    \caption{Demonstration of the hierarchical scheme, the first panel shows the reconstructed SNR time-series without any averaging and a single trigger (SNR $\geq 4$). In the middle panel is the averaged SNR series obtained from binned averaging of $w=16$ samples (shown in green in the first panel), and the red region corresponds to the bin containing the original trigger. The last panel shows a finer reconstruction performed only for the bin having average SNR greater than the first stage threshold.}
    \label{zoomed}
\end{figure}

The reconstruction of SNR time-series for every sample and each template requires large computational costs. Since we are only interested in triggers exceeding a certain threshold, it is better not to reconstruct the complete SNR times-series, rather only in the vicinity of the triggers. We propose a two-step hierarchical scheme for reconstruction, which performs a coarse reconstruction, and then a finer reconstruction around the triggers obtained in the first stage. In the first stage, we consider fixed non-overlapping bins of $w$ samples. Then the outputs from each basis vector $\beta_k$ are averaged in the bins referred as the averaged time-series 

\begin{align}
    \beta^{avg}_k [t_i] = \sum \limits_{j=0}^{w-1} \beta_k [t_{i\times w + j}] / w,
\end{align}

where $i = \{0,...,N/w-1\}$ corresponds to different bins, and $j = \{0,..,w-1\}$ goes over the samples inside each bin. The bin index $i$ can be considered as a sample point in the shortened averaged time-series. By linearly combining the $\beta^{avg}_k$ with the decomposition coefficients $c_{k, \zeta}$ similar to Eq. (\ref{Beta_eqn}), results into an averaged SNR times-series for the template $h_{\zeta}$. We identify the first stage triggers and their corresponding bins having the average SNR above a first stage threshold $\rho_{\text{I}}$. In the next step we perform a finer reconstruction for every sample inside all the triggering bins. Triggers from the second stage which are above a second stage threshold $\rho_{\text{II}}$ are referred as the final triggers. In the Fig. \ref{zoomed}, we show the hierarchical reconstruction of SNR time-series around a trigger.
 
Our aim is to minimize the total costs for the hierarchical method without losing any sensitivity. The sensitivity is determined by computing the fraction of triggers recovered at a given SNR, or conversely, the SNR threshold at which all the triggers are recovered. We compute the sensitivity using the latter approach -- by comparing the hierarchical distribution of the triggers with the original distribution. The parameters $w, \rho_{\text{I}}$ are fine-tuned under the constraint of reaching a fixed target SNR ($\rho_{target}$) to optimize the total costs. We give a detailed description to compute $\rho_{target}$ in the later subsection \ref{Sec:Cost-estimation}.

\subsection{Fast First Stage Filtering using Templates}\label{sec:fft_first_stage}

To reduce the first stage costs the average SNR can be obtained by the faster template method which is mathematically equivalent to reduced basis filtering. Using the template method, the average SNR time-series can be obtained by binned averaging of the integrand in the Eq. (\ref{IFFT}) and then performing an IFFT of shorter length. To demonstrate the averaging process mathematically, we consider a single bin $b$ with $w$ samples, and compute the averaged SNR for the same, which is denoted by $\Braket{\rho_{\theta}(t)}_b$.  
The $w$ time samples in the bin are represented by $t = wb + r$, where $r \in [0, w-1]$. We then write the discretized version of the Eq. (\ref{IFFT}) averaging the SNR time-series over $w$ samples 

\begin{align}
    \Braket{\rho_{\zeta}(t)}_b & = \dfrac{1}{Nw} \sum \limits_{r=0}^{w-1} 4 \Delta f \sum \limits_{f=0}^{N-1}\dfrac{\tilde{s}[f]\tilde{h}_{\zeta}^*[f]}{S_n[f]}e^{2\pi i f(wb+r)/N}. \label{eq:avg_fft1}
\end{align}
Now, breaking the summation over $f \in [0,N-1]$ into a double sum by expressing $f = lN/w + f'$, we can rewrite the single summation of any function $\tilde{g}(f)$ in the Fourier domain as
\begin{align}
    \label{eq:double-sum}
    \sum \limits_{f = 0}^{N-1} \tilde{g}[f] = \sum \limits_{f'=0}^{N/w-1} \sum \limits_{l=0}^{w-1} \tilde{g}\Big[l\frac{N}{w} + f'\Big],
\end{align}
where $l \in [0,w-1]$ and $f' \in [0, N/w-1]$. This modifies the Eq.~(\ref{eq:avg_fft1}) to 
\begin{align}
\label{eq:avg_fft2}
\begin{split}
    \Braket{\rho_{\zeta}(t)}_b &= \dfrac{4 w\Delta f}{N}\sum \limits_{f'=0}^{N/w-1} e^{2 \pi if' \frac{w}{N}b} 
    \\
    &\times \underbrace{\dfrac{1}{w^2}\sum \limits_{l=0}^{w-1} \dfrac{\tilde{s}[l\frac{N}{w} + f']\tilde{h}_{\zeta}^*[l\frac{N}{w} + f']}{S_n[l\frac{N}{w} + f']} \sum \limits_{r = 0}^{w-1} e^{2\pi i(l\frac{N}{w} + f')\frac{r}{N}}}_{=\Omega(f')}
    \\
    & = \dfrac{4w\Delta f}{N}\sum \limits_{f'=0}^{N/w-1} e^{2\pi i f'\frac{w}{N}b} \Omega(f').
\end{split}
\end{align}
To simplify Eq. (\ref{eq:avg_fft2}), we introduce a new frequency series $\Omega(f')$ which represents the binned average of the PSD weighted correlation of the data and template. The second summation term in $\Omega(f')$ has a closed form solution and can be computed analytically. We notice that the last line of Eq. (\ref{eq:avg_fft2}) has a similar form to Eq. (\ref{IFFT}) but, the integrand replaced with $\Omega(f')$ having only $N/w$ number of samples. Suggesting that the average SNR time-series can be obtained by an IFFT of $\Omega(f')$, where one can think of $b$ as the new time equivalent variable which ranges between $[0, N/w-1]$. Therefore, it is possible to perform the first stage using basis or templates, however, the second stage filtering needs to be done with the basis.

\subsection{Implementation}
We now discuss the implementation of our method which is divided into two parts -- 1) preparation and 2) matched filtering. The preparation stage is implemented partially on GPUs, whereas matched filtering is performed entirely on the GPUs. We use several Nvidia GV100 GPUs, each having memory of 32 GB as well as several RTX 2070 Super each with 8 GB of memory. Our code is written in C language and uses various optimized libraries for different purposes. Operations on the GPU are performed using CUDA \cite{CUDA} an application programming interface by Nvidia.

In the preparation stage, we perform PCA on the template bank to obtain the reduced basis and the respective decomposition coefficients. Matrix multiplications on the GPUs in this stage are performed using the \textit{cuBLAS} library from CUDA. We begin by computing the covariance matrix in several parts in parallel using cuBLAS. Afterward we combine all the parts to obtain the final matrix $\textbf{C}$. The Lanczos algorithm for EVD of $\textbf{C}$ is implemented on CPUs using the SLEPc \cite{SLEPc} and PETSc \cite{PETSc} libraries. We obtain the decomposition coefficients matrix $\mathbf{D}$ by multiplying the matrices $\textbf{T}$ and $\textbf{P}$. The preparation stage is computed in advance and performed only once. Results from this stage -- the basis and the decomposition coefficients, are stored on hard drives for the later matched filtering stage. To reduce the input/output (IO) bandwidth, we compress the PCA results before writing them on the hard drive.  

In the next stage, we read the output from the previous stage to match filter data using our hierarchical scheme. To reduce the time-intensive memory transfers between CPU and GPU, we load the matrices $\mathbf{P}$ and $\mathbf{D}$ at once onto the GPUs. We divide the data into several smaller segments such that we can optimally utilize the memory of the GPUs while filtering each segment in parallel. We use data segments of $128$s and a sampling rate of 2048 Hz. Each data segment has $N = 128 \times 2048$ samples which overlap with $N/2$ samples from the previous segment. We employ the cuFFT library from CUDA to perform the FFTs in this stage. Using cuFFT we perform FFT for a batch of data segments in parallel. In the next step, we interpolate the basis vectors and multiply them with $\tilde{s}$ along with $S_n(f)^{-1/2}$. Afterward we perform in-place batched IFFTs to obtain the filtering output from the basis. The in-place technique saves GPU memory by recycling the allocated input memory to write the output.  

To perform the hierarchical reconstruction we first average the output from basis to obtain the $\pmb{\beta}^{avg} $ matrix. The first stage reconstruction is done using cuBLAS by multiplying $\pmb{\beta}^{avg}$ and $\textbf{D}$, which outputs the average SNR time-series. Next, we use a dedicated function on the GPU to find triggers with average SNR above $\rho_{\text{I}}$. Once the triggers are identified, we store their bin indices along with their corresponding average SNRs. Since these triggers are not contiguous in memory, further reconstruction of the triggering bins cannot be performed by simple matrix multiplication. Therefore, instead of using the optimized cuBLAS library, we use a custom-built function on the GPU to perform the second reconstruction.

\subsection{Cost Estimation}\label{Sec:Cost-estimation}
In this subsection, we estimate the floating-point operations required by the two different matched filtering schemes. The purpose of estimating the number of operations is to get a rough idea of the scaling relations involved for the total filtering costs. Moreover, it will allow us to estimate the improvement in performance due to the proposed hierarchical scheme. In both methods, we split the data into blocks of $N$ samples, having an overlap of $N/2$ samples with the previous block. This is generally done to avoid corrupt SNR samples at the start and end of a data segment \cite{find_chirp}. Hence, filtering a single block results in $N/2$ unique SNR time-samples. Most operations involve complex numbers unless otherwise specified. Throughout the cost estimation, we consider 6 operations for multiplication and 2 operations for the addition of two complex numbers. To estimate the costs for the FFTs we consider a split-radix method \cite{FFT-op}.

We first estimate the costs for the template method which will be our baseline comparison. The first step is to compute the forward real-to-half-complex FFT of the data with $N$ samples, which requires $3/2N\text{log}N$ operations per block \cite{FFT-op}. Computing the integrand of the matched filtering Eq.(\ref{MF_def}) for $T$ templates requires $6NT$ operations. Finally, an inverse complex-to-complex FFT is required to obtain the SNR time-series for each template, and this attributes to $5TN\text{log}N$ operations, where each IFFT requires $5N\log N$ operations. In total the template method for a single block requires $N\text{log}N(3/2+5T) + 6TN$ operations. Usually, the number of templates is huge ($T \gg 1$), so we can neglect the cost for the forward FFT of data. Therefore, the total floating-point operations $z_{basic}$ for filtering $N/2$ data samples with the template method can be approximated to
\begin{align}
\label{FFT_operations}
     z_{basic} = NT(5\log N + 6).
\end{align}

Now, we estimate the costs for the two stage hierarchical filtering. As shown in the subsections \ref{sec:first_stage} and \ref{sec:fft_first_stage}, the first stage can be performed either by using the basis or the templates. We evaluate the first stage costs by considering the faster template method described in the section \ref{sec:fft_first_stage}. Starting with the forward FFT of the data which needs $3/2N\text{log}N$ operations. Afterward the weighted correlation of the matched filter in Eq. (\ref{eq:avg_fft2}) is then obtained for every template in $6NT$ multiplicative operations. Then, we perform binned averaging of the correlations to get reduced frequency series $\Omega(f')$ of size $N/w$ for each template, and this requires $2NT/w$ operations. For every template, we obtain the average time series by computing the IFFT of $\Omega(f')$ in $5NT/w\log (N/w)$ operations. Hence, the number of floating-point operations (neglecting the forward FFT) required for the first stage is 
\begin{align}
    \begin{split}
    \label{first_stage_cost}
        z_{first} = NT\Bigg(\dfrac{5}{w}\log(\frac{N}{w}) + 6 + \frac{2}{w}\Bigg).
    \end{split}
\end{align}

In the next stage, we compute a finer reconstruction of $w$ points around each first stage trigger. The costs for the second stage are calculated in terms of the number of first stage triggers. We denote the number of first stage triggers for a single template by $f(w, \rho_{\text{I}})$. We assume that the number of triggers do not vary for different templates, and thus, can be obtained from a single test template. This assumption is justified for the templates in the vicinity of the test template as the number of triggers would be roughly the same. In addition, since $f(w, \rho_{\text{I}})$ decreases rapidly at higher SNRs, the error in the total costs due to our assumption is negligible. Since the fast first stage does not involve computing the basis outputs $\beta_k$, we evaluate them in the second stage in $Np(5\log(N) + 6)$ operations. Using the above assumption, the second stage requires $z_{second} = 4pwf(w,\rho_{\text{I}})T + Np(5\log(N)+6)$ operations. Summing up the costs from both the stages, the total floating-point operations required for the hierarchical method are
\begin{align}
    \begin{split}
    \label{final_cost}
    z_{total} = & NT\Bigg(\dfrac{5}{w} \log(\frac{N}{w}) + 6 + \frac{2}{w}\Bigg) \\
         & + \Big( 4pw f(w, \rho_{\text{I}})T + Np(5\log(N) + 6)\Big).
    \end{split}
\end{align}

The final costs in Eq. (\ref{final_cost}) are obtained in terms of two nuisance parameters $w$ and $\rho_{\text{I}}$. An appropriate choice of the first stage threshold $\rho_{\text{I}}$ is important in determining the background trigger distribution. This is to not miss any potential triggers in the first stage, because only triggers that are followed till the second stage are accounted in the background estimation. To ensure that we recover all the triggers using a reliable first stage threshold in the two stage filtering scheme, we compare the hierarchical distribution of second stage triggers against the trigger distribution from the flat scheme. The idea is to identify a target SNR ($\rho_{target}$) as a function of ($w, \rho_{\text{I}}$), such that the hierarchical scheme recovers $99\%$ of the total triggers from the flat scheme. Considering a certain first stage configuration given by specific values of ($w, \rho_{\text{I}}$), we denote the number of final triggers above $\rho_{\text{II}}$ as $n_{final}(\rho_{\text{II}})$. Similarly, using the same threshold $\rho_{\text{II}}$, we denote triggers from the flat scheme as $n_{flat}(\rho_{\text{II}})$.
The target SNR $\rho_{target}$ is then defined as
\begin{align}
\label{target_snr_criteria}
    \rho_{target}(w, \rho_{\text{I}}) = \Big(\min (\rho_{\text{II}}) \Big| n_{final}(\rho_{\text{II}}) \geq 0.99 n_{flat}(\rho_{\text{II}}) \Big).
\end{align}

\subsection{Low pass filter interpretation for the first stage}
There are many ways to coarse filter the data to produce a reduced SNR time-series; one of such methods is using a low pass filter combined with decimation. Our proposed method of averaging the SNR time series (described in section \ref{sec:fft_first_stage}) resembles qualitatively a low pass filter, however it takes into account the entire frequency range as seen in Eq. (\ref{eq:double-sum}). 

\begin{figure}[H]
    \centering
    \includegraphics[width=\linewidth, height = 7cm]{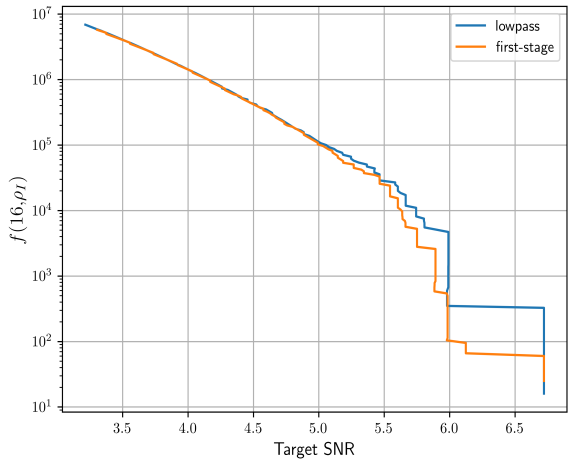}
    \caption{Comparing two different coarse filtering methods -- low pass (blue) and first stage (orange). We show the number of first stage triggers $f(16, \rho_{I})$ for simulated Gaussian noise as a function of target SNR. The two methods have broadly the same performance except for SNRs $> 5$ where the low pass method registers up to 10 $\times$ more false alarms. }
    \label{lowpass_vs_first}
\end{figure}

We test the performance of a low pass and our first stage filtering methods by filtering simulated Gaussian noise. A metric for comparison could be the number of false alarms produced for a given target SNR. Since the target SNR depends on the thresholding criterion we obtain the thresholds respectively for each method. Using the Eq. (\ref{target_snr_criteria}), we obtain the number of first stage triggers $f(16, \rho_{\text{I}})$ as a function of the target SNR as shown in the Fig. \ref{lowpass_vs_first}. We observe, that both methods have very similar but not identical performance. Both methods produce the same number of triggers at SNRs $\leq 5$, but for SNRs $> 5$ the low pass method triggers more false alarms than the first stage -- we observe up to 10 times more false alarms. This indicates that because the first stage filter preserves the high frequency content, it has slightly better sensitivity over a low pass filter.

\section{Accuracy and Performance Analysis of the hierarchical method}\label{Sec:Results}

In the previous section, we demonstrated our method and estimated the required costs for filtering. In this section, we present the accuracy of the hierarchical filtering results and assess the reduction in the required number of operations w.r.t. the baseline -- template method. Furthermore, we measure the gain in performance by implementing matched filtering on GPUs relative to the established CPU implementations.

We use the sub-region (shown in Fig. \ref{Bank} in orange) to demonstrate our method. This sub-region covers the parameter ranges $M \in [5.72, 12.05]$ and $q \in [1.0, 11.05]$. Using a tolerance of $10^{-5}$ as per Eq. (\ref{SNR_loss}), we obtain $p = 254$ for this sub-region. We want to estimate a conservative reduction in the total costs that scale linearly with $p$. Following this reason, we choose the mentioned sub-region as it corresponds closely to the average $\Braket{p}$.

Current offline and online searches \cite{pycbc_offline, pycbc-live} generally use SNR threshold of $\sim 4-5$ for the single detector triggers. However, searches involving a large number of templates are affected by increased background due to noise triggers \cite{ianharry_precession, Nitz_ecc_search}, and thus, higher SNR thresholds are used to detect events at a constant false alarm rate. These kinds of search scenarios also happen to be the case where cost-efficient algorithms are necessary. Therefore, in this work, we target SNR thresholds of 5 and above. 

\subsection{Accuracy of the SNR}
\begin{figure}[H]
    \centering
    \includegraphics[width=\linewidth, height = 7cm]{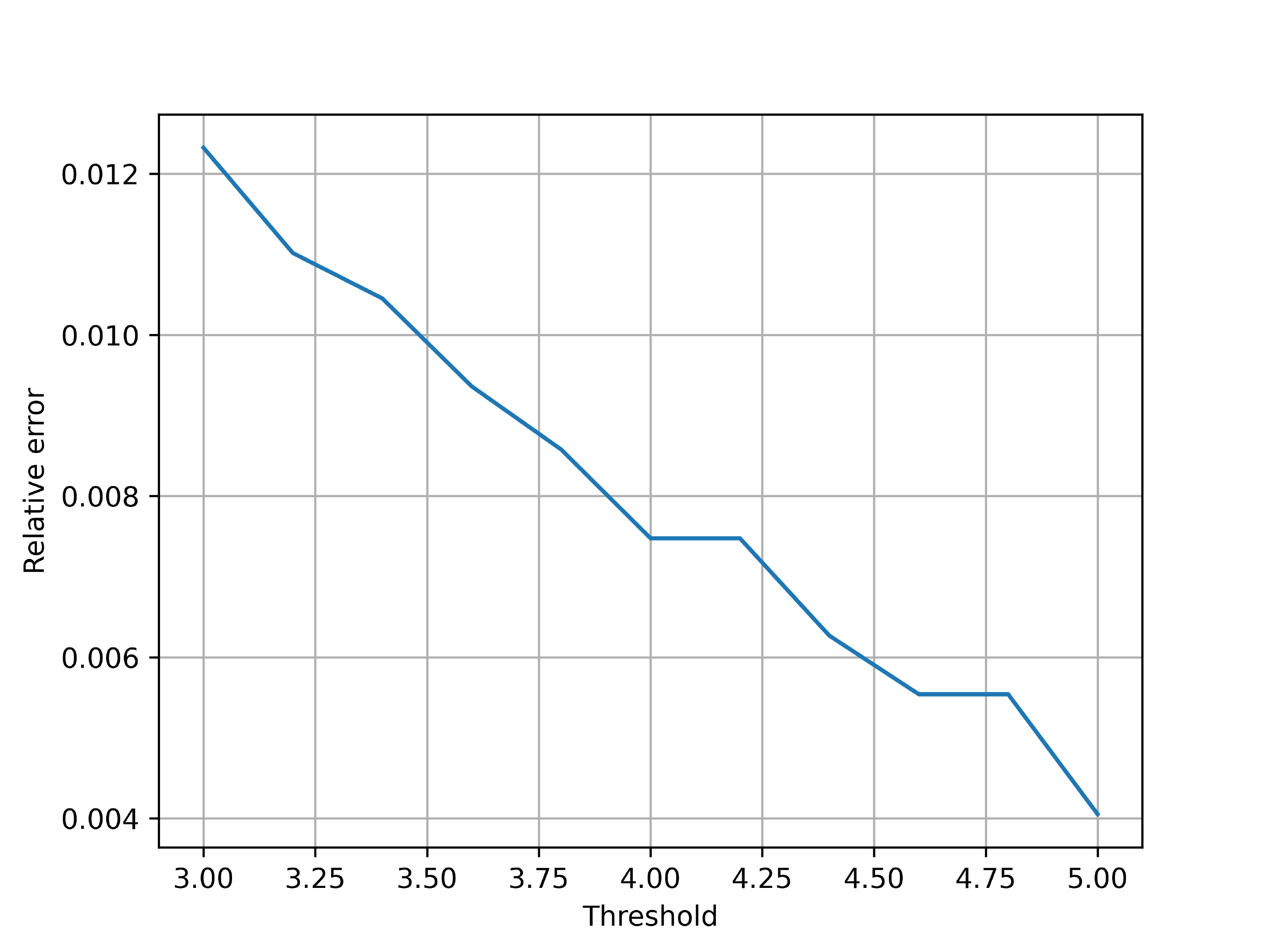}
    \caption{Maximum relative error between the original and reconstructed SNR values as a function of SNR threshold. The incurred error scales inversely with the SNR thresholds and is smaller than the error due to the discreteness of the templates for relevant SNR thresholds.}
    \label{relative_error}
\end{figure}

We expect two primary contributions to the SNR loss in our method - \textit{truncation of the number of eigenvalues} and \textit{interpolation of the basis}. Error due to truncating the eigenvalues is translated as the SNR loss via the Eq. (\ref{SNR_loss}), and is regulated by choosing an appropriate number of basis vectors. The loss due to linear interpolation of the basis is quantified in terms of the mismatch between the interpolated and fully sampled templates. We evaluate the total loss by computing the relative error in the SNR time-series obtained using our method and the template method. We filter simulated colored Gaussian noise from the PSD to acquire the SNR time-series. Comparing triggers from every template, we note the maximum relative error in the SNR values and plot it against the SNR thresholds as shown in Fig. \ref{relative_error}.

Based on our sensitivity requirements we are interested only in SNR = 5 and above, and from Fig. \ref{relative_error} we observe the relative error to be $\sim 0.4\%$ for SNR = 5.0. This amount of error can be tolerated because the observed loss is less than the error due to mismatch of the templates, which is up to $3\%$. We also notice that the relative error decreases even further with higher SNR thresholds. Therefore, it is justified that our method successfully recovers the SNR values for search scenarios requiring SNR thresholds $\geq 5$.

\subsection{Comparing Performance with Template Based Matched Filtering}
Our hierarchical method is characterised by two parameters, the averaging bin size $w$, and the first stage threshold $\rho_{\text{I}}$. For testing the hierarchical scheme, we use four different averaging bin size $w \times dt$ with $w \in [2, 4, 8, 16]$, and various different values of $\rho_{\text{I}}$. We drop the $dt$ factor when referring $w$, to suggest the reader that we simply average over $w$ samples. We test our scheme using noise generated from the aLIGO PSD sampled at 2048 Hz. We also check our method for a population of BBH signals within the case study sub-region. Finally, we compare the estimated costs required by the hierarchical method against the template method. We use PyCBC software library \cite{pycbc-software} to generate simulated data containing Gaussian noise and to perform injections.

\subsubsection{Number of required operations}
Using the estimates in section \ref{Sec:Cost-estimation}, we compare the number of operations required by the hierarchical scheme with the template method. We begin by testing only Gaussian noise generated using the aLIGO PSD. For this purpose, we generate a total of $\sim 7.4$ days of data using different seeds sampled at 2048 Hz. For filtering purposes, the simulated data is then divided in smaller segments of 128s with an overlap of 64s from the previous segment.  

\begin{figure}[h]
    \centering
    \includegraphics[width=\linewidth, height = 7.5cm]{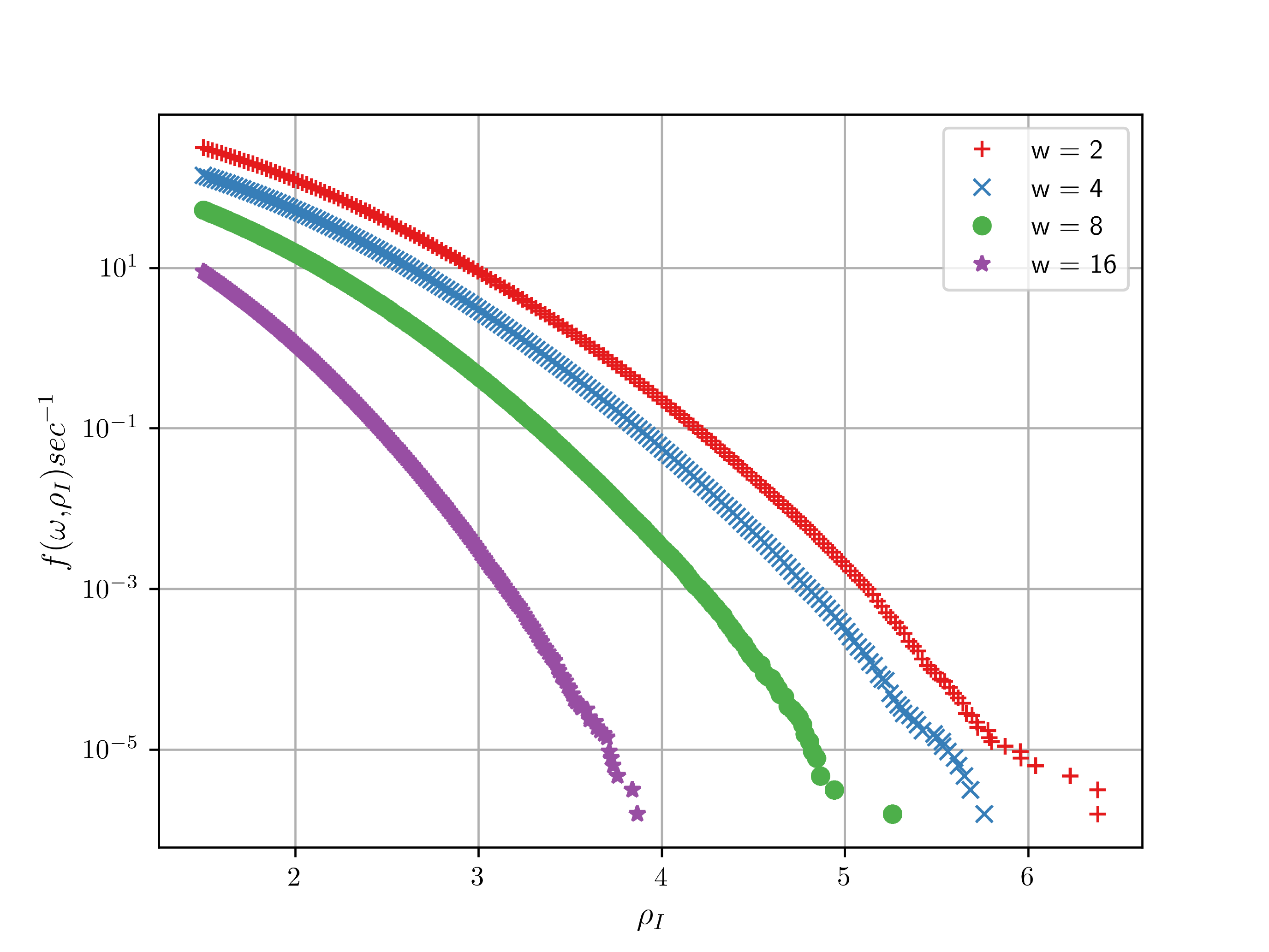}
    \caption{The number of first stage triggers $f(\omega, \rho_{\text{I}})$ per second as a function of $\rho_{\text{I}}$ for different averaging bin sizes $w$ (shown in the legend). As $w$ is increased, $\rho_{\text{I}}$ must be decreased to recover the same number of triggers $f$.}
    \label{Triggers_vs_cutoff}
\end{figure}

To estimate the total costs for the hierarchical method Eq. (\ref{final_cost}), we first determine the number of first stage triggers $f(w, \rho_{\text{I}})$ by varying the hierarchical parameters. We neglect the variation of $f(w, \rho_{\text{I}})$ for different templates for reasons discussed previously in section \ref{Sec:Cost-estimation}. $f(w, \rho_{\text{I}})$ is obtained simply by iterating over different values of the first stage threshold $\rho_{\text{I}}$ for different (fixed) $w$ and by counting the total first stage triggers above the same threshold. The observed $f(w, \rho_{\text{I}})$ w.r.t. $\rho_{\text{I}}$ for different values of $w$ is shown in Fig. \ref{Triggers_vs_cutoff}.

It is seen from Fig. \ref{Triggers_vs_cutoff} that the number of first stage triggers decreases with increasing $\rho_{\text{I}}$, as expected. But more interestingly, we notice that fewer triggers are recovered when the averaging is done over more samples for the same $\rho_{\text{I}}$, i.e. bigger $w$.  Therefore, to maintain the same sensitivity or the number of final triggers while $w$ increases, the first stage threshold must be lowered. 

\begin{figure}[h]
    \centering
    \includegraphics[width=\linewidth, height = 7.5cm]{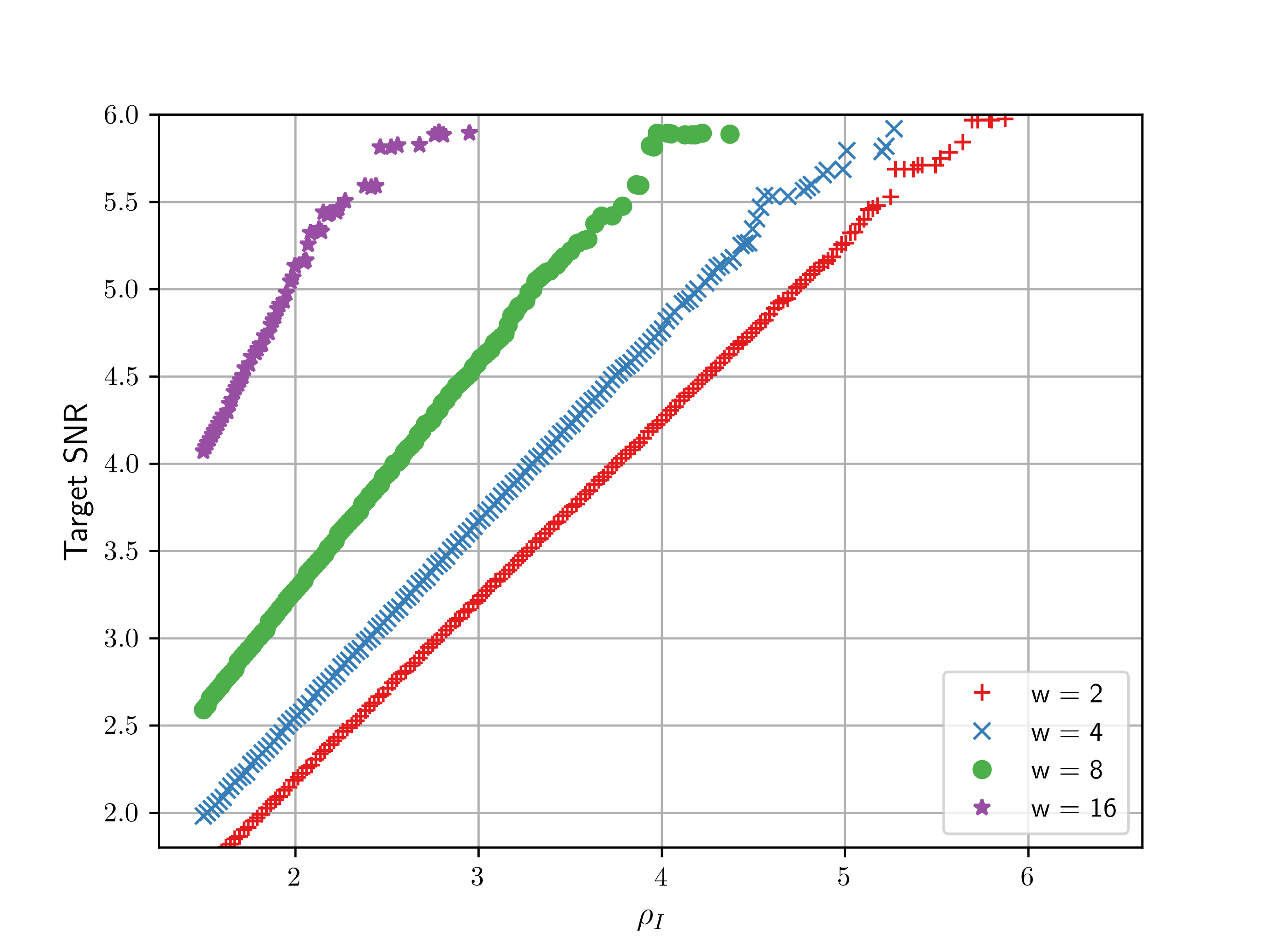}
    \caption{ The plot shows different combinations of $w$ and $\rho_{\text{I}}$ to achieve a specific target SNR $\rho_{target}$ without losing any sensitivity. The choice of $w$ and $\rho_{\text{I}}$ determines the total cost of the hierarchical method for the corresponding $\rho_{target}$. 
   }
    \label{cutoff_vs_target}
\end{figure}

A particular combination of the parameters $(w, \rho_{\text{I}})$ determines a specific target SNR $\rho_{target}$ for the hierarchical method without any loss in sensitivity as discussed in section \ref{Sec:Cost-estimation}. The target SNR is computed for a fixed $w$ and different values of $\rho_{\text{I}}$, by iterating over different values of second stage SNR $\rho_{\text{II}}$ using Eq. (\ref{target_snr_criteria}). In Fig. \ref{cutoff_vs_target} we plot the relationship between $\rho_{\text{I}}$ and $\rho_{target}$ for different $w$ . Using the results obtained in Fig. \ref{cutoff_vs_target}, we can choose various combinations of $w$ and $\rho_{\text{I}}$ to reach a desired $\rho_{target}$. We notice from the plot that to reach the same $\rho_{target}$, bigger $w$ requires a lower value of $\rho_{\text{I}}$. This information combined with the previous plot suggests that for a specific $\rho_{target}$, choosing a bigger $w$ leads to more number of first stage triggers.

\begin{figure*}[ht]
    \centering
    \includegraphics[width=\linewidth, height = 13cm]{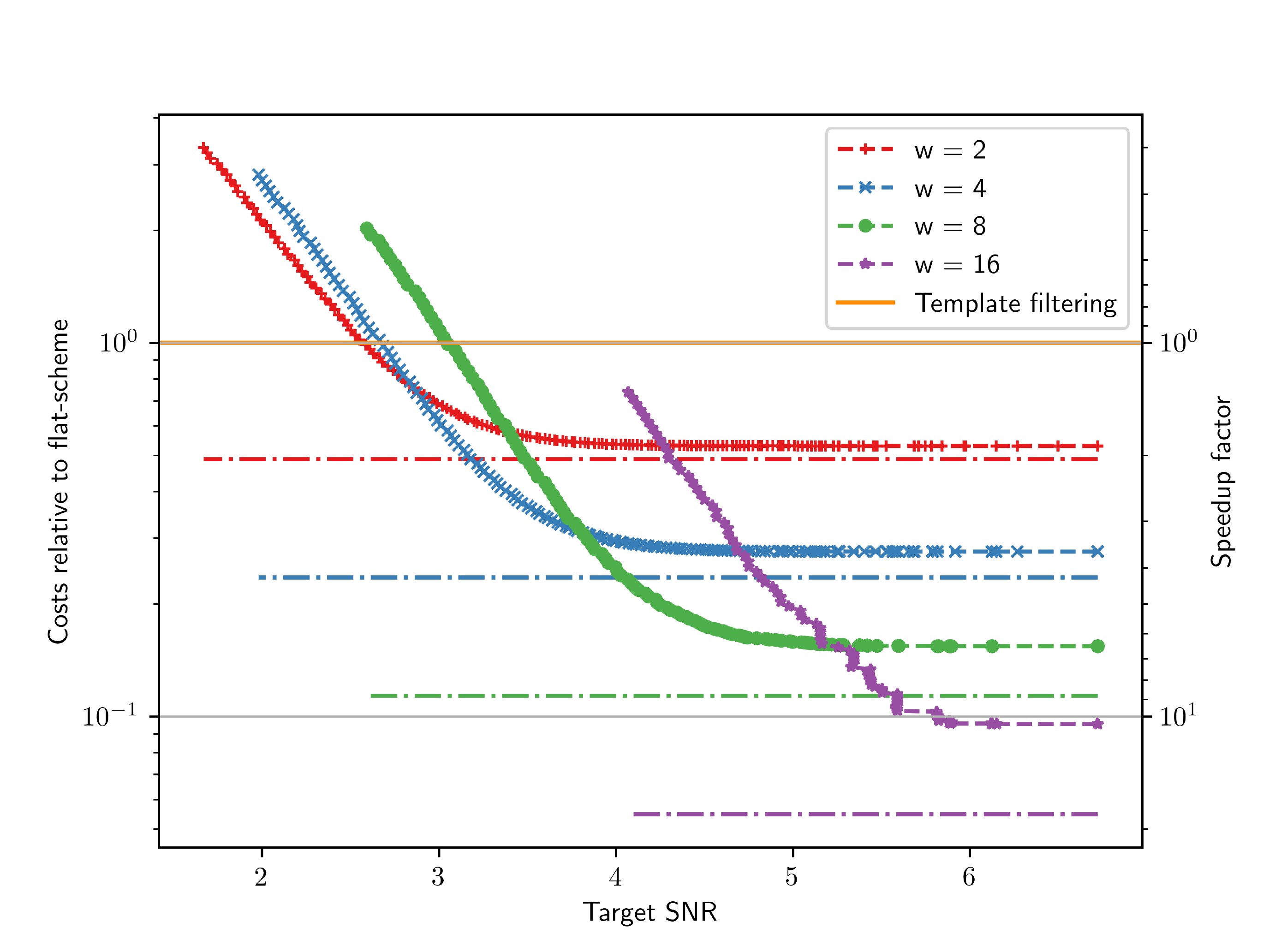}
    \caption{The computational cost (in terms of FLOP) for hierarchical filtering $\sim 7.4$ days of simulated data containing only Gaussian noise against 6250 templates. We show the costs for different (fixed) $w$ which are normalized by the direct template filtering cost (orange line) versus the target SNR. The first stage costs are indicated by dot-dash lines, and the lines with markers correspond to the total costs. On the right, the vertical axis shows the relative speed-up factor (log scale) compared to the template method. The best speed up for a desired $\rho_{target}$ is achieved by choosing $w$ accordingly from this plot.}
    \label{Costs_stack}
\end{figure*}

Now, we combine the results from Fig. \ref{Triggers_vs_cutoff} and Fig. \ref{cutoff_vs_target} to estimate the final costs Eq. (\ref{final_cost}) in terms of $\rho_{target}$ as shown in the Fig. \ref{Costs_stack}. These costs are normalized by the costs required by the template method Eq. (\ref{FFT_operations}) (shown by the horizontal orange line). To demonstrate the contributions in the total costs from the individual stages separately, we plot the first stage and the total costs together. The first stage costs are constant w.r.t. $\rho_{target}$ and scale inversely with $w$. On the other hand, due to the rapid increase of the first stage triggers $f(w, \rho_{\text{I}})$ at low SNRs, the second stage costs become dominant. It is also inferred from the figure that the second stage costs are more for larger $w$ at a constant $\rho_{target}$. We notice from the Fig. $\ref{Costs_stack}$, the total costs converge to the first stage costs at higher SNRs, and hence, infer that the first stage leads to the dominant costs for the hierarchical matched filtering.  

From Fig. \ref{Costs_stack} we observe a reduction in total costs compared to the baseline for all choices of $w$ complying to our desired target of $\rho_{target} \geq 4.0$. We notice for $\rho_{target} = 5.0$, the setting with $w=8$ achieves a relative speed up factor of 6, which corresponds to a reduction of $\sim 83\%$ in the total costs. It is observed that the hierarchical method performs better with increasing SNR thresholds. Our method achieves the best computational gain of $\sim 10$ times which is equivalent to a reduction of $\sim 90\%$ in the total costs, for $\rho_{target} = 6.0$ using $w = 16$. We have not tested the scheme for higher values of $w$, but by extrapolating the obtained results we may infer that the hierarchical method might perform better with $w > 16$ for even higher SNR thresholds.

\begin{figure}[H]
    \centering
    \includegraphics[width=\linewidth, height = 7cm]{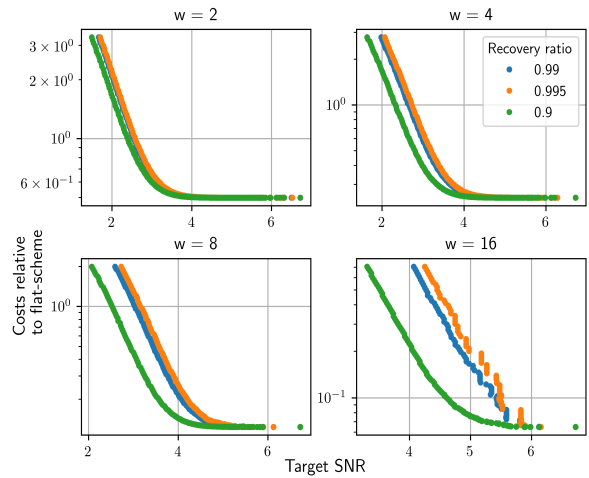}
    \caption{Varying the recovery ratio used in the target SNR. We use three values of the recovery ratio 0.99 (blue), 0.995 (orange) and 0.9 (green) to test its impact on the total costs. We show the total relative costs versus the target SNR for four different window sizes. We notice the total costs are not sensitive for small changes in the recovery ratio, but can be further fine-tuned using this ratio as a parameter.}
    \label{Fig:recovery_ratio}
\end{figure}

While estimating the target SNR we use a fixed value of the recovery ratio = $99\%$ (see Eq. (\ref{target_snr_criteria})). To understand the impact of varying the recovery ratio, we compare the total costs for three values of the recovery ratio -- 0.995, 0.99 and 0.9. We plot the total costs as a function of target SNR for all averaging window sizes in the Fig. \ref{Fig:recovery_ratio}. From the plot we observe that there is no significant change in the total costs at SNRs $\geq 6$ and for smaller windows. But we notice a further reduction in costs for larger windows, especially at low SNR values, in exchange for the reduced accuracy. We infer from Fig. 8 a general trend for the total cost that is proportional to the recovery ratio; the curve translates to left or right, respectively. Moreover, the shift increases with the window size. This suggests that our results are not sensitive to small changes in the recovery ratio when close to unity. Depending on the search requirements, the recovery ratio is another parameter for further tuning performance.

To ensure that higher SNR triggers are recovered using only $\rho_{\text{I}}$, and also to sanity check our method for recovering CBC signals. We test our scheme for a population of 1000 CBC injections randomly generated within the case study region. For every injection, we create separate strains of 128 sec and place signals randomly into simulated data containing only colored Gaussian noise. The signals correspond to SNRs ranging between [4.5, 30]. We use two test cases of $\rho_{target} =$ 4.0 and 6.0 using $w=16$. Using the appropriate first stage cutoff $\rho_{\text{I}}$ from Fig. \ref{cutoff_vs_target} we recover all the injections that are above the respective test cases of SNR threshold. Therefore, we verify the reliability of our method to recover CBC signals using only $\rho_{\text{I}}$ even at higher SNRs.

We now make a few key remarks. The total costs of the hierarchical method are dominated by the first stage at higher SNR thresholds, which can be reduced by choosing a bigger $w$. On the other hand, the second stage costs are dominant at lower SNRs which grow as $w$ increases. As seen in Fig. \ref{Costs_stack}, for different SNRs we obtain different optimal window lengths, e.g. for $\rho_{target} = 5.0$ the best setting is when $w=8$. Depending upon the threshold criterion required for the search the optimal choice of $w$ can be chosen based on Fig. \ref{Costs_stack}. In case, the search demands higher SNR thresholds than shown in the plot, a larger $w$ may be preferred to further reduce the costs.

\subsubsection{Observed Performance}
In this section, we measure the performance of matched filtering implementations on GPUs to estimate a realistic improvement compared to the established search pipelines. For this purpose, we use a widely quoted performance metric -- the \textit{throughput} of a search method, used for determining the number of templates analyzed in real-time. Consider a data segment of $N$ secs filtered against $T$ templates and the filtering process takes $t$ seconds, then the throughput would be $NT/t$ templates processed in real-time. Once again, we filter simulated colored Gaussian noise sampled at 2048 Hz for 64 seconds to evaluate this metric.

We benchmark the template method implemented using the optimized \textit{cuFFT} library from CUDA. The template method in-situ took roughly 100ms to filter 64 secs of data against 6250 templates per GPU. Therefore, achieving an in-situ performance of $4000 \times 10^3$ templates processed in real-time on a single Nvidia GV100. We want to remark that the second stage reconstruction is not optimized and thus, we could not benchmark the hierarchical scheme to its full potential. Considering the costs from Fig. \ref{Costs_stack}, we estimate the expected peak performance of a completely optimized hierarchical method, which suggests that the hierarchical implementation may require only 12ms to perform the cuFFT equivalent filtering. Hence, we expect an increase of roughly an order of magnitude in the throughput  (second row in Table \ref{table:commercial_metrics}) if the second stage is fully optimized. Work is in progress for optimizing the second stage.  

We now compare performances using previously quoted numbers from currently used search pipelines -- PyCBC live \cite{pycbc-live} and PyCBC offline \cite{pycbc_offline}, as shown in Table \ref{table:commercial_metrics}. The PyCBC search methods are implemented on multiple CPU cores, whereas ours are on multiple GPUs. Depending upon the search method the throughput is standardized by templates processed in real-time per core (GPU) for PyCBC (hierarchical) implementation. We notice from the first column in Table \ref{table:commercial_metrics} that using the latest GPUs gives an enormous improvement in the throughput. However, the PyCBC numbers are not quoted from an up-to-date hardware implementation and hence, a fair comparison might require the latest hardware. 

Furthermore, we also present commercially motivated metrics to benchmark the performance, measuring the cost and energy efficiency of hardware while filtering. These metrics are computed by normalizing the throughput by the total cost or the energy consumption of the hardware respectively. The two metrics for the different search schemes are listed in Table \ref{table:commercial_metrics}. 
\begin{table*}[ht]
    \begin{center}
    \resizebox{9.2cm}{2.8cm}{%
    \begin{tabular}{cccc}
        \hline
        \rule{0pt}{1\normalbaselineskip} &  &\\
         Method & Throughput & Throughput/ & Throughput/ \\
                & & Euro & W \\
         \hline \hline
         \rule{0pt}{1\normalbaselineskip} & &\\
         cuFFT(in-situ) & 4000 x $10^3$ & 400 & 14 x $10^3$\\
         \hline 
         \rule{0pt}{1\normalbaselineskip} & &\\
            Hierarchical scheme (expected) & 2300 x $10^4$ & 2300 & 82 x $10^3$\\
         \hline 
         \rule{0pt}{1\normalbaselineskip} & &\\
         PyCBC live  & 6300 & 17 & 31\\
         \hline 
         \rule{0pt}{1\normalbaselineskip} & &\\
         PyCBC offline  & 12,000 & 32 & 60\\
         \hline\\
    \end{tabular}
    }
    \end{center} 
    \caption{Comparing different implementations of the matched filtering schemes on GPUs (first and second row) with the established PyCBC schemes on CPUs (third and fourth row). In the second column we show the throughput for the respective methods, and in the third and fourth columns are the throughput per euro and per watt of power consumption respectively. The expected peak performance of the hierarchical method is estimated for SNR = 5.0 in the second row.}  \label{table:commercial_metrics}
\end{table*}

It is apparent from Table \ref{table:commercial_metrics}, that due to GPU's ability to perform tasks in a highly parallelized fashion, GPUs can analyze $\sim 10$x templates more than CPUs for the same costs. We also notice that considering the same power consumption a single GPU is equivalent to $\sim 10^3$ CPU cores for analyzing templates at a given instant. These metrics suggest that GPUs are highly energy and cost-efficient in performing matched filtering, which motivates their application in CBC searches. In addition, proper implementation of our hierarchical method will allow further improvement in the efficiency of the hardware and will help reduce the time required to perform extensive offline searches.

\section{Discussion and future prospects}\label{Discussion_and_future_prospects}

In this work, we have demonstrated using simulated data containing Gaussian noise an efficient way of matched filtering. We filter using a reduced basis and employ a new hierarchical method to reduce the reconstructions costs. Compared to the template based filtering, our method is $\sim 10\times$ faster than the template based filtering methods without losing sensitivity at a threshold SNR = 6, and $\sim 6 \times$ for SNR = 5. The gain in performance increases with higher SNR thresholds and is currently estimated for a specific region of the parameter space. Our method is successful in recovering the original flat search background, and thus, does not compromise the significance of detected candidates with SNR above the SNR threshold.

We demonstrate the advantages of implementing matched filtering methods on the latest GPUs. We compare the throughput of GPU implementation of matched filtering with the CPU implementation of current methods. Benchmarking the in-situ performance of template method implementation on GPUs, we observe a performance gain of 2-3 orders in magnitude compared to the PyCBC search pipelines. Our results indicate a significant improvement in performance, which may motivate the development of a fully optimized second stage reconstruction.  In addition, we present two new metrics to compare the performance of the matched filtering implementation on different hardware. Analyzing these metrics suggests that GPUs are more cost and energy-efficient in performing matched filtering than CPUs. Hence, the utilization of GPUs is encouraged for current or future searches.

A possible avenue to improve the described method would be to find better ways of performing the first stage and a faster implementation of the second stage. In this work, we use a constant sampling rate for matched filtering. Multi-rate sampling can be implemented to further improve the performance of the hierarchical method for cases where latency is a strong requirement or the duration of signals is significantly longer than those tested here. 

In the near future, detectors will become more sensitive and thus the cost-effective hierarchical method proposed here can be useful for exploring sub-solar regimes or searching for low frequency long duration signals. Our method might play a role in reducing the computational costs for the future 3G detectors where the template bank size can be at least an order of magnitude larger \cite{ianharry_precession, ET-templates} than the current CBC banks. Furthermore, this method can also be employed in new regions of the parameter space to perform computationally intensive searches for sources exhibiting precession or eccentricity.

\acknowledgments
We thank Badri Krishnan, Kipp , and Tom Dent for the valuable discussions. We also thank Marlin Sch{\"a}fer and Yifan Wang for their comments on the manuscript. We acknowledge the Max Planck Gesellschaft and the Atlas cluster computing team at Albert-Einstein Institute (AEI) Hannover for support.

\newpage
\bibliography{references}

\end{document}